\magnification1200 \baselineskip14pt

\centerline{\bf Desperately Seeking Superstrings?}
\smallskip
\centerline{by Paul Ginsparg and Sheldon Glashow}
\smallskip
\centerline{Physics Today, May 1986}
\bigskip

Why is the smart money all tied up in strings?
Why is so much theoretical capital expended upon the properties of
supersymmetric systems of quantum strings propagating in ten-dimensional
space-time?
The good news is that superstring theory may have the right stuff to explain
the ``low-energy phenomena'' of high-energy physics and gravity as well. In the
context of possible quantum theories of gravity, each of the few currently
known superstring theories may even be unique, finite and self-consistent. In
principle a superstring theory ordains what particles exist and what properties
they have, using no arbitrary or adjustable parameters. The bad news is that
years of intense effort by dozens of the best and the brightest have yielded
not one verifiable prediction, nor should any soon be expected. Called ``the
new physics'' by its promoters, it is not even known to encompass the old and
established standard model.

In lieu of the traditional confrontation between theory and experiment,
superstring theorists pursue an inner harmony where elegance, uniqueness and
beauty define truth. The theory depends for its existence upon magical
coincidences, miraculous cancellations and relations among seemingly unrelated
(and possibly undiscovered) fields of mathematics. Are these properties reasons
to accept the reality of superstrings? Do mathematics and aesthetics supplant
and transcend mere experiment? Will the mundane phenomenological problems that
we know as physics simply come out in the wash in some distant tomorrow? Is
further experimental endeavor not only difficult and expensive but unnecessary
and irrelevant? Contemplation of superstrings may evolve into an activity as
remote from conventional particle physics as particle physics is from
chemistry, to be conducted at schools of divinity by future equivalents of
medieval theologians. For the first time since the Dark Ages, we can see how
our noble search may end, with faith replacing science once again. Superstring
sentiments eerily recall ``arguments from design'' for the existence of a
supreme being. Was it only in jest that a leading string theorist suggested
that ``superstrings may prove as successful as God, Who has after all lasted
for millennia and is still invoked in some quarters as a Theory of Nature''?

The trouble began with quantum chromodynamics, an integral part of the standard
model that underlies the quark structure of nucleons and the nuclear force
itself. QCD is not merely a theory but, within a certain context, {\it the\/}
theory of the strong force: It offers a complete description of nuclear and
particle physics at accessible energies. While most questions are
computationally too difficult for QCD to answer fully, it has had many
qualitative (and a few quantitative) confirmations. That QCD is almost
certainly ``correct'' suggests and affirms the belief that elegance and
uniqueness --- in this case, reinforced by experiment -- are criteria for
truth.

No observed phenomenon disagrees with or demands structure beyond the standard
model. No internal contradictions and few loose ends remain, but there are some
vexing puzzles: Why is the gauge group what it is, and what provides the
mechanism for its breakdown? Why are there three families of fundamental
fermions when one would seem to suffice? Aren't 17 basic particles and 17
tunable parameters too many? What about a quantum theory of gravity? Quantum
field theory doesn't address these questions, and one can understand its
greatest past triumphs without necessarily regarding it as fundamental. Field
theory is clearly not the end of the story, so something smaller and better is
needed: Enter the superstring.

The trouble is that most of superstring physics lies up at the Planck mass ---
about $10^{19}$ GeV -- and it is a long and treacherous road down to where we
can see the light of day. A naive comparison of length scales suggests that to
calculate the electron mass from superstrings would be a trillion times more
difficult than to explain human behavior in terms of atomic physics.
Superstring theory, unless it allows an approximation scheme for yielding
useful and testable physical information, might be the sort of thing that
Wolfgang Pauli would have said is ``not even wrong.'' It would continue to
attract newcomers to the field simply because it is the only obvious
alternative to explaining why certain detectors light up like video games near
the end of every funding cycle.

In the old days we moved up in energy step by step, seeing smaller and smaller
structures. Observations led to theories or models that suggested further
experiments. The going is getting rougher; Colliders are inordinately
expensive, detectors have grown immense, and interesting collisions are rare.
Not even a politically popular ``Superstring Detection Initiative'' with a
catchy name like ``String Wars'' could get us to energies where superstrings
are relevant. We are stuck with a gap of 16 orders of magnitude between
theoretical strings and observable particles, unbridgeable by any currently
envisioned experiment. Conventional grand unified theories, which also depend
on a remote fundamental energy scale (albeit one extrapolated upward from known
phenomena rather than downward from abstract principle), retain the grand
virtue that, at least in their simplest form, they were predictive enough to be
excluded --- by our failure to observe proton decay.

How tempting is the top-down approach! How satisfying and economical to explain
everything in one bold stroke of our aesthetic, mathematical or intuitive
sensibilities, thus displaying the power of positive thinking without requiring
tedious experimentation! But {\it a priori\/} arguments have deluded us from
ancient Greece on. Without benefit of the experimental provocation that led to
Maxwell's equations and, inevitably, to the special theory of relativity, great
philosophers pondering for millennia failed even to suspect the basic
kinematical structure of space-time. Pure thought could not anticipate the
quantum. And even had Albert Einstein succeeded in the quest that consumed the
latter half of his life, somehow finding a framework for unifying
electromagnetism and gravity, we would by now have discarded his theory in the
light of experimental data to which he had no access. He had to fail, simply
because he didn't know enough physics. Today we can't exclude the possibility
that micro-unicorns might be thriving at a length scale of $10^{-18}$ cm.
Einstein's path, the search for unification now, is likely to remain fruitless.

Having a potentially plausible candidate ``theory of everything'' does
dramatically alter the situation. But we who are haunted by the lingering
suspicion that superstrings, despite all the hoopla, may be correct are likely
to remain haunted for the foreseeable future. Only a continued influx of
experimental ideas and data can allow the paths from top and bottom to meet.
The theory of everything may come in its time, but not until we are certain
that Nature has exhausted her bag of performable tricks.

\bye